\documentclass[10pt,letterpaper]{IEEEtran}
\usepackage{amsmath,amssymb,amsfonts}
\usepackage{gensymb}%for the "degree" symbol
\usepackage{balance}

\usepackage{graphicx}
\usepackage{cite}
\usepackage{amsthm}
\usepackage{stackengine}
\usepackage[dvipsnames,usenames]{color}
\usepackage[colorlinks, linkcolor=Blue, filecolor =Red,citecolor=RoyalPurple]{hyperref}
\usepackage{scalerel}
\usepackage{nopageno}
\usepackage{ifthen}
\IEEEoverridecommandlockouts
\usepackage[lmargin=0.76in,rmargin=0.76in,tmargin=0.73in,bmargin=0.78in]{geometry}
\newtheorem{comment}{Comment}

\newtheorem{lem}{Lemma}

\def\delequal{\mathrel{\ensurestackMath{\stackunder[1pt]{=}{\scriptstyle\triangledown}}}}

\newtheorem{assumption}{Assumption}
\newtheorem{cor}{Corollary}
\newtheorem{argu}{Argument}

\def\R{\mathbb{R}}

\def\Expect{{\sf E}}
\def\eqdef{:=}

\newcommand{\version}{arxiv}
\newcommand{\markedManu}{MARKED}
\newboolean{showArxiv}
\newboolean{showArxivAlt}

\ifthenelse{\equal{\version}{journal}}
{
	\setboolean{showArxiv}{false} 
	\setboolean{showArxivAlt}{true}
}
{
}

\ifthenelse{\equal{\version}{arxiv}}
{
	\setboolean{showArxiv}{true} 
	\setboolean{showArxivAlt}{false}
}
{
}

\ifthenelse{\equal{\markedManu}{MARKED}}
{

}
{
}

\ifthenelse{\equal{\markedManu}{nonMARKED}}
{

}
{
}

\def\optProbName{planning}
\def\textAgg{\text{agg}}
\def\textOn{\ensuremath{\hbox{on}}}
\def\textOff{\ensuremath{\hbox{off}}}
\def\textBA{\ensuremath{BA}}

\def\textSet{\ensuremath{\hbox{set}}}

\def\Ptotmax{\ensuremath{P}^{\text{agg}}}
\def\stuckOnKJ{\ensuremath{S^{\textOn,j}_{k}}}
\def\stuckOffKJ{\ensuremath{S^{\textOff,j}_{k}}}
\def\stuckOnK{\ensuremath{s^{\textOn}_{k}}}
\def\stuckOffK{\ensuremath{s^{\textOff}_{k}}}

\def\stuckOn{\ensuremath{s^{\textOn}}}
\def\stuckOff{\ensuremath{s^{\textOff}}}

\def\switchOff{\ensuremath{f^{\textOff}}}
\def\flipOn{\ensuremath{f^{\textOn}}}
\def\flipOff{\ensuremath{f^{\textOff}}}

\def\tauTCL{\ensuremath{\tau_{\scaleto{tcl}{4pt}}}}
\def\tauBA{\ensuremath{\tau}}
\def\rd#1{{\color{red}{#1}}}

\newlength{\noteWidth}
\long\def\notes#1{\ifinner
          {\footnotesize #1}
          \else
          \marginpar{\parbox[t]{\noteWidth}{\raggedright\footnotesize #1}}
      \fi\typeout{#1}}
\def\pb#1{\notes{pb: \rd{#1}      }}

\usepackage{xparse}

\ExplSyntaxOn
\cs_new:Nn \pb_prop_gset_bykeys:Nn
 {
  \prop_clear:N \l__pb_temp_prop
  \keys_set:nn { pb/propbykey } { #2 }
  \prop_gset_eq:NN #1 \l__pb_temp_prop
 }
\cs_generate_variant:Nn \pb_prop_gset_bykeys:Nn { c }

\keys_define:nn { pb/propbykey }
 {
  unknown .code:n = \prop_put:Nxn \l__pb_temp_prop { \l_keys_key_tl } { #1 }
 }

\cs_generate_variant:Nn \prop_put:Nnn { Nx }

\NewDocumentCommand{\setupcollaborator}{mm}
 {% #1 = identifier string, #2 = set of key-value pairs
  \prop_new:c { g_collaborator_#1_prop }
  \pb_prop_gset_bykeys:cn { g_collaborator_#1_prop } { #2 }
 }

\NewDocumentCommand{\selectcollaborator}{m}
 {
  \prop_map_inline:cn { g_collaborator_#1_prop }
   {
    \tl_set:cn { ##1 } { ##2 }
   }
 }
\ExplSyntaxOff

\setupcollaborator{PBmacbookair}
 {% pb, macbook air
  NAME=Prabir Mac Laptop,
  PBbibPATH=/Users/pbarooah/Dropbox/PBbib,
  DiCEbibPATH=/Users/pbarooah/Dropbox/Dropbox/dicelab_bib,
  ACbibPATH=/Users/pbarooah/Dropbox/Dropbox/austin_bib,
  JBbibPATH=/Users/pbarooah/Dropbox/Dropbox/bibs,
  figPATH=/Users/pbarooah/Dropbox/papers/18_loads/figures/,
  REFRESHMENT=coffee,
 }
\setupcollaborator{AClinux}
 {% AC is AC using the linux machine at work
  NAME=AC at lab,
  figPATH=\string~/Dropbox/,
  DiCEbibPATH=../../../bibFiles/dicelab_bib,
  ACbibPATH=../../../bibFiles/austin_bib,
  JBbibPATH=../../../bibFiles/bibs,
  REFRESHMENT=ice cream,
 }
\selectcollaborator{PBmacbookair} 
\begin{document}	

\title{\vspace{6.4mm}Aggregate capacity of TCLs with cycling constraints}

\author{
	\IEEEauthorblockN{Austin R. Coffman\IEEEauthorrefmark{1}$^,$\IEEEauthorrefmark{2}, Neil Cammardella\IEEEauthorrefmark{1}, Prabir Barooah\IEEEauthorrefmark{1}, and Sean Meyn\IEEEauthorrefmark{1}}
	\thanks{\IEEEauthorrefmark{1} University of Florida} 
	\thanks{\IEEEauthorrefmark{2} corresponding author, email: bubbaroney@ufl.edu.}
	\thanks{AC and PB are with the Dept. of Mechanical and Aerospace Engineering, University of Florida, Gainesville, FL 32601, USA. NC and SM are with the department of Electrical and Computer Engineering, University of Florida, Gainesville, Fl 32601, USA. The research reported here has been partially supported by the NSF through award 1646229 (CPS-ECCS). %Additionally, we thank Ana Bu\v{s}i\'{c} for her contributions to our earlier joint work on TCLs that benefited this paper.
        }
	\vspace{-0.75cm}
}

%\author{\IEEEauthorblockN{Austin Coffman}
%	\IEEEauthorblockA{School of Electrical and\\Computer Engineering\\
%		Georgia Institute of Technology\\
%		Atlanta, Georgia 30332--0250\\
%		Email: http://www.michaelshell.org/contact.html}
%	\and
%	\IEEEauthorblockN{Ana Bu\v{s}i\'{c}}
%	\IEEEauthorblockA{Twentieth Century Fox\\
%		Springfield, USA\\
%		Email: homer@thesimpsons.com}
%	\and
%	\IEEEauthorblockN{Prabir Barooah}
%	\IEEEauthorblockA{Starfleet Academy\\
%		San Francisco, California 96678-2391\\
%		Telephone: (800) 555--1212\\
%		Fax: (888) 555--1212}}

%%%%%%%%%%%%%%%%%%%%%%%%%%%%%%

\maketitle
\thispagestyle{empty}
\begin{abstract}
Thermostatically Controlled Loads (TCLs) such as air conditioners and water heaters typically maintain their temperature within a preset range using on/off actuation. These types of loads are inherently flexible: many different power consumption trajectories exist that can keep the temperature within range. Decades of research has shown that flexible loads can provide valuable grid services.

Quantifying the power and energy capacities of a collection of TCLs is a well-studied problem. However, most works focus on temperature constraints. In this work, we present a characterization of the capacity of a collection of TCLs that considers not only temperature, but also cycling and energy constraints. The characterization leads to a set of convex constraints. A grid operator can use this characterization to compute a feasible power consumption trajectory for an ensemble of TCLs that comes closest to what the operator needs to maintain demand-supply balance. Unlike prior attempts at capacity characterizations incorporating cycling constraints, our results are independent of the algorithm used to coordinate the TCLs.
\end{abstract}

\section{Introduction}\label{sec:intro}
Currently, power balance in power grids is maintained mostly through supply-side actions, i.e., generators are ramped up and down to meet demand, resulting in negative economic and environmental impacts. These negative impacts motivate an active area of research: controlling flexible loads to provide grid support. Some examples of flexible loads that are suitable for grid support are Thermostatically Controlled Loads (TCLs)~\cite{callaway2011achieving,ChenDistributedIMA:2017,CoffmanStudyHPB:2018,matkoccal:2013,zhang2013aggregated,liu2019trajectory}, HVAC systems in commercial buildings~\cite{haokowlinbarmey:2013}, heat pumps~\cite{LeeGridJESBC:2020}, and electric pumps for irrigation~\cite{AghajFarm:2019} and pool cleaning~\cite{chebusmey17a}. 

Flexible loads can alter their power consumption without violating Quality of Service (QoS) constraints. A grid operator or balancing authority (BA) can request an ensemble of flexible loads that they consume more or less power with respect to a baseline. Baseline refers to the power consumption that would have occurred without the BA's interference. From the perspective of the BA, an increase (or decrease) of power consumption is identical to the charging (or discharging) of a battery. Due to this similarity, these resources are often termed \textit{Virtual Energy Storage} (VES)~\cite{bar:2019}. However, VES is cheaper than grid-scale batteries~\cite{cammardella2018energy}.

An extensive literature exists on reference tracking by collections of TCLs~\cite{matkoccal:2013,ChenDistributedIMA:2017,zhang2013aggregated,CoffmanVESBuildSys:2018,liushi:2016}. The BA computes a reference signal (say, in MW) for a collection of loads, and a coordination algorithm has to ensure that the total power consumption deviation of the collection (from the baseline) tracks this reference. However, in order for a BA to design a \textit{feasible} reference signal, the capacity of the collection of TCLs must be known. Though no formal definition of capacity exists, in this context capacity denotes limits on aggregate power consumption deviation due to QoS constraints at the individual loads. For TCLs, there are at least three QoS constraints: (i) temperature, (ii) cycling rate, and (iii) total energy consumption. If a BA designs a reference signal with an incomplete notion of capacity, the BA must accept poor tracking or the TCL users must accept QoS violations. In both scenarios the long-term outlook is grim: either the BA views TCLs as an unreliable resource, or the TCL users view the BA as an authoritative monarch with unrealistic expectations. 

A significant amount of research has focused on characterizing demand flexibility capacity of TCLs~\cite{hao_aggregate:2015,ZhaoGeometricTPS:2017,ZirasNewPSCC:2018,SanandajiRampTSE:2016,ChengHierarchicalEPES:2019,CoffmanAggregate:CDC:19}. % deferrable load capacity \cite{rooz:2018}, and the capacity for flexible loads~\cite{ChenAggregateTSG:2020,MullerAggregationTSG:2019}.
The impact of enforcing QoS constraints on the loads on reference tracking performance, when the reference is planned without considering capacity, has also been investigated~\cite{ChenMarkovianThesis:2016,CoffmanVESBuildSys:2018}. Early work in this area only accounted for temperature constraints \cite{hao_aggregate:2015,ZhaoGeometricTPS:2017}. More recent work has included cycling constraints~\cite{ZirasNewPSCC:2018,SanandajiRampTSE:2016,ChengHierarchicalEPES:2019,WangFlexibilityTSG:2020}, but their capacity characterization is limited to specific coordination algorithms, and furthermore not suitable for planning a feasible reference. 

%In \cite{ZirasNewPSCC:2018}, cycling constraints are considered, but reference planning is not done. Furthermore, the aggregate power constraint is algorithm-dependent. Similar to~\cite{ZirasNewPSCC:2018} the cycling constraint in~\cite{ChengHierarchicalEPES:2019} is also algorithm-dependent. In~\cite{SanandajiRampTSE:2016}, a centralized approach is used to handle aggregate capacity for TCLs with cycling constraints. The authors provide a ramp rate constraint on the power deviation reference signal to account for the cycling constraint. However, using the constraints for reference planning requires a simulation of TCLs, which renders the capacity characterization algorithm-dependent. In summary, the surveyed works suffer from a number of limitations: they do not consider all QoS constraints, are algorithm-dependent, or are not convenient for BA-level reference planning. 

%One advantage of this paradigm is that the capacity characterized through it allows for efficient and coordination algorithm independent computation of reference signals for ensembles of TCLs. However, in so far, there has been no extension of the advantages of this paradigm for when TCLs have cycling constraints. }

In this work we develop a capacity characterization that accounts for three QoS constraints at the individual TCLs: temperature, cycling, and energy. The key novelty of our characterization is its ability to account for cycling constraints. The characterization is independent of the algorithm used to control the ensemble of TCLs. Two, the capacity characterization can be used by a BA to compute the reference for an ensemble of TCLs by solving an optimization problem that is always \emph{feasible} and \emph{convex}. Together, these features ensure that the reference signal so planned can be tracked with any well-designed coordination algorithm that respects the three QoS constraints of each TCL. 

The effectiveness of our capacity characterization is demonstrated in simulation experiments by comparing reference tracking performance with two distinct references: one planned with our method and another planned with a method that is representative of prior work that does not incorporate cycling constraints. The capacity characterization we develop requires making an approximate homogeneity assumption, but the numerical results show the method is robust to those assumptions.

Our work starts with the paradigm introduced in~\cite{hao_aggregate:2015} of constructing an ensemble battery model. The discrete nature of a TCL's power consumption was ignored in~\cite{hao_aggregate:2015} to develop an average temperature like quantity for the collection. Extending this framework to incorporate cycling constraints is challenging due to the on-off nature of TCLs control that leads to an integer valued constraints on number of cycles. We address this challenge by transferring the constraints into quantities that involve fraction of loads at various states, thereby eliminating integer-valued constraints. 

This work extends our recent work~\cite{CoffmanAggregate:CDC:19} in the following two ways. The reference planning problem developed here is convex and independent of the coordination algorithm used, whereas the one in~\cite{CoffmanAggregate:CDC:19} is non-convex and only applicable to a specific coordination algorithm. A preliminary version of this paper is published in~\cite{coffmanFlexibility_ACC:2020}. The results in~\cite{coffmanFlexibility_ACC:2020} is for strictly homogeneous loads, which is extended to a class of heterogeneous loads here. A more extensive numerical investigation is included here to demonstrate the effectiveness of the method, especially with heterogeneous loads.

%An ensemble of TCLs are coordinated with a priority stack controller (a modified version of the one developed in~\cite{hao_aggregate:2015}, so as to enforce device cycling constraints), to track the computed reference. We offer a comparison of reference planning and tracking when the reference signal is planned using the aggregate capacity constraints of~\cite{hao_aggregate:2015}, which do not include information on individual TCLs cycling and energy QoS. The results of the comparison confirm the need to include all relevant individual TCL QoS requirements in reference planning.

The paper proceeds as follows: Section~\ref{sec:invTCL} contains descriptions of individual TCL behavior, Section~\ref{sec:aggQuans} contains descriptions of aggregate TCL behavior, and Section~\ref{sec:propMethod} contains the derived aggregate capacity constraints. In Section \ref{sec:propRefGen}, the proposed reference planning method is described. Lastly, Section~\ref{sec:numExp} reports the results of numerical experiments.
% These aggregate constraints do not account for the individual TCLs cycling constraint. It is shown in simulations that the priority stack controller is unable to track the reference when it is generated without information on the individual TCLs cycling constraints.
%Need to cite this~\cite{deng2015survey}.
%Some past smart grid comm papers:~\cite{guo2018fast}, \cite{ashraf2018logarithmic}, \cite{bahrami2018autonomous}, and~\cite{khezeli2016data}. 

\ifx 0
\section{Needs of the Grid} \label{sec:gridNeeds}
It is envisioned that collections of TCLs will be a resource for the BA to help eliminate demand supply mismatch. The BA enlists TCLs as a resource by first obtaining a signal, we term $r_k^{BA}$, that represents some notion of supply/demand mismatch in a given geographical area. This signal should then be projected onto a set that represents the capacity of a collection of TCLs. This step is necessary as there is no reason that $r_k^{BA}$ should by itself be feasible for a collection of TCLs. The process of projecting $r_k^{BA}$ onto the constraint set of a collection of TCLs is deemed the \textit{reference \optProbName} problem.

The feasible portion as obtained from the \textit{reference \optProbName} problem is $r_k$, and this is the signal the BA expects a collection of TCLs to track. By design this reference signal will always be a zero mean deviation signal so that the TCLs are not asked to act as generators. To track $r_k$, the collection of TCLs need to consume $r_k$ more/less than their baseline power consumption, which we denote by $\bar{P}$. %Baseline power consumption refers to the power consumption that would have occurred without the interference from the BA, which we denote by $\bar{P}$ (for the collection). 
% In 
%Figure~\ref{fig:bpaRefPlot} an example of $r_k $ and $r_k^{BA}$ is shown, here $r_k^{BA}$ is BPA's balancing reserves signal. The signal $r_k$ is generated by \algoName, which is described in the rest of the paper.

%\begin{figure}
%	\centering
%	\includegraphics[width=1\columnwidth, height = 0.35\columnwidth]{bpaRefPlot.pdf}
%	\caption{An example single of $r_k^{BA}$ and the feasible portion for a collection of TCLs, $r_k$.}
%	\label{fig:bpaRefPlot}
%\end{figure}
\fi

\section{The Individual TCL} \label{sec:invTCL}
An on/off TCL is any device that turns on or off to maintain a temperature within a preset temperature deadband. Time is discrete, with a sampling period $T_s$, and is denoted by the index $k$. There are $N$ TCLs, indexed by $j=1,\dots,N$. The temperature of the $j$-th TCL at discrete time instant $k$ is denoted by $\theta_k^j$ and its on/off status during the continuous time interval $[kT_s, (k+1)T_s)$ is denoted by $m^j_k$ (=1 if on and 0 if off). We denote the \emph{rated electrical power consumption} of the $j$-th TCL, the power consumed by it when on, by the constant $P^j$. 

\subsection{Modeling a  TCL's temperature} \label{sec:detCont}
As in much of prior work~\cite{matkoccal:2013,CoffmanVESBuildSys:2018} temporal evolution of the temperature $\theta_k^j$ of the $j$-th TCL is modeled in discrete time as a linear difference equation
\begin{align}\label{eq:dynTCL}
\theta_{k+1}^j = a^j\theta_{k}^j + (1-a^j) \left(\theta_{k}^a - R^j_{th}m_k^j\eta^jP^j \right)
\end{align}
with $a^j \triangleq \exp\left(\frac{-T_s}{R^j_{th}C^j_{th}}\right)$, where $R^j_{th}$ and $C^j_{th}$ represent the thermal resistance to ambient temperature $\theta_{k}^a$ and thermal capacitance, respectively. For an air conditioner (AC) providing cooling, the term $\eta^j P^j$ is the thermal power rejected to the ambient by the TCL $j$ when it is on, and $\eta^j$ is its Coefficient of Performance (COP).

%Throughout the paper we consider a TCL providing cooling (such as an air conditioner) and therefore assume that $\theta_k^a > \theta^j_{\textSet} + \delta^j$ for each TCL $j$ and all $k$, where $\delta^j$ is the deadband of the thermostat. 

For later use we now define the \emph{analytical baseline demand} of the $j$-th TCL, $\bar{P}^j_k$: it is the electrical power demand needed to maintain $\theta_k^j$ at the setpoint, $\theta^j_{\textSet}$ for all $k$. Because eventually we are interested in aggregate quantities over the whole collection, it is common to ignore the binary nature of power consumption at this stage; see, e.g.,~\cite{hao_aggregate:2015}. The qualifier ``analytical'' is used to emphasize that this is a quantity introduced for analysis: such a demand cannot be observed for a single TCL.  The analytical baseline demand can be computed by finding the value of $P^j_k$ in~\eqref{eq:dynTCL} that ensures an equilibrium of~\eqref{eq:dynTCL}, with $\theta^j_{k+1} = \theta^j_{k}=\theta^j_{\textSet}$ for all $k$. It follows from straightforward calculations that
\begin{align} \label{eq:invBase}
	{P}^{j,b}_k = \frac{\theta_{k}^a - \theta^j_{\textSet}}{\eta^j R^j_{th}}.
\end{align}
The analytical baseline is a time varying quantity since the ambient temperature $\theta_k^a$ is time-varying.
\ifx 0
A related quantity commonly used in the literature is the thermal energy deviation of the $j^{th}$ TCL~\cite{hao_aggregate:2015}, denoted: 
\begin{align} \label{eq:invThemEng}
	z_{k}^j \triangleq \frac{C^j_{th}}{\eta^j}(\theta_{k}^j - \theta^j_{\textSet}).
\end{align}
The dynamics for thermal energy are obtained by substituting the definitions for $z_k^j$ and $z_{k+1}^j$ into~\eqref{eq:dynTCL},
\begin{align} \label{eq:dynInvThemEng}
	&z_{k+1}^j = a^jz_k^j - b^j\left(m_k^jP^j - \frac{\theta_{k}^a - \theta^j_{\textSet}}{\eta^j R^j_{th}}\right), \\ \label{eq:bDef}
	&b^j = (1-a^j)C^j_{th}R^j_{th}.
\end{align} 
The quantity in parenthesis in the RHS of~\eqref{eq:dynInvThemEng} is the power deviation for the $j^{th}$ TCL from the baseline. 
\fi

\subsection{QoS constraints for a TCL}
The quality of service constraints (QoS) for the $j^{th}$ TCL are:
\begin{align} \label{eq:invTz}
	\text{QoS 1:}& \quad \left|\theta_{k}^j - \theta_{\textSet}^j\right| \leq \delta^j, \quad \forall \ k,\\ \label{eq:invCycle}
	\text{QoS 2:}& \quad \sum_{i=0}^{\tauTCL^j-1}\left|m_{k-i}^j-m_{k-1-i}^j\right| \leq 1, \quad \forall \ k, \\ \label{eq:invEng}
	\text{QoS 3:}& \quad T_s\left|\sum_{k=0}^{H_b}\left(m_k^jP^j - \hat{P}_k^j\right)\right| \leq \tilde{E}^j. 
\end{align}
The first constraint says that TCL $j$'s temperature must be kept within $\pm \delta^j$ of the setpoint $\theta_{\hbox{set}}$, where $\delta^j$ is a predetermined constant. For later reference, we note that the full width temperature deadband is denoted as $\Delta \triangleq 2\delta$. The second is the cycling constraint; it says that the device can only flip -- from either on to off or from off to on - once within a specified period $\tauTCL^j$. The third is a constraint on the energy consumed over the billing horizon $H_b^j$: it says the total energy consumed by the TCL over a horizon $H_b^j$ cannot deviate from its (analytical) baseline by more than a specified amount, $\tilde{E}^j$ ($>0$). Just like the temperature deadband, the parameters $H_b^j,\tilde{E}^j$ are design choices that depend on the $j$-th consumer's preference. For instance, if the consumer wishes that the energy use over 30 days do not vary by more than 10\% of a baseline energy use of $1000$ kWh, then $H_b^j= \frac{60}{5}\times 24 \times 30 = 8640$ (for a 5-minute sampling period) and $\tilde{E}^j=100$ kWh.

The set of TCL-specific parameters that appear in~\eqref{eq:dynTCL},\eqref{eq:invTz}-\eqref{eq:invEng} is $Q_s^j \triangleq \{\theta_{\textSet}, \delta, \tauTCL, \tilde{E}, H_b,R_{th},C_{th},\eta,a,P\}^j$. A subset of these specifies the QoS constraints of the consumer while the remaining describe mechanical/thermal properties of the hardware. 

For later use, we now define variables to describe a TCL's state of flipping from on to off (or vice versa) state, and the state of being stuck in the on (or off) state. The ``flip on'' or ``flip off'' variables are defined as
\begin{align}
\text{\big(Flip on\big)}& \quad	F^{\textOn,j}_{k-1} \triangleq \begin{cases}
	1, & \text{if} \ (m^{j}_{k}-m^{j}_{k-1}) = 1.\\
	0, & \text{otherwise.}
	\end{cases} \\	
\text{\big(Flip off\big)}& \quad	F^{\textOff,j}_{k-1} \triangleq \begin{cases}
	1, & \text{if} \ (m^{j}_{k-1}-m^{j}_{k}) = 1.\\
	0, & \text{otherwise.}
	\end{cases}
\end{align}
\ifx 0
An on or off flip can occur because of two events: (i) the TCL flips its on/off status to maintain the temperature QoS~\eqref{eq:invTz} or (ii) the TCL flips for the purpose of providing VES:
\begin{align}
	F^{\textOn,j}_{k-1} &= V_{k-1}^{\textOn,j} + D_{k-1}^{\textOn,j}, \\
	F^{\textOff,j}_{k-1} &= V_{k-1}^{\textOff,j} + D_{k-1}^{\textOff,j}.
\end{align}
The quantity $V_{k}^{\textOn,j}$ (respectively, $V_{k}^{\textOff,j}$) represents the off to on flip at $k$ that occurs to provide VES (respectively, off to on flip). The quantity $D_{k}^{\textOn,j}$ (respectively, $D_{k}^{\textOff,j}$) represents a flip that occurs to maintain the temperature within its deadband, i.e., QoS~\eqref{eq:invTz}.\pb{remove now?} \fi
We say a TCL is \textit{stuck on} (respectively, \textit{stuck off}) at time $k$ if it is off (respectively, on) at that time and has changed mode once in the past $\tauTCL$ time instants, so that it is unable to switch mode at the current time $k$. We define the stuck on and off state as \stuckOnKJ \ and \stuckOffKJ \ :
\begin{align} \nonumber
	\stuckOnKJ \triangleq \begin{cases}
	1, \ \text{if} \ \sum_{i=0}^{\tauTCL^j-1}\left|m^{j}_{k-i}-m^{j}_{k-1-i}\right| = 1, \ m_k^{j} = 1.\\
	0, \ \text{otherwise.}
	\end{cases} \\ \nonumber
	\stuckOffKJ \triangleq \begin{cases}
	1, \ \text{if} \ \sum_{i=0}^{\tauTCL^j-1}\left|m^{j}_{k-i}-m^{j}_{k-1-i}\right| = 1, \ m_k^{j} = 0.\\
	0, \ \text{otherwise}.
	\end{cases}
\end{align} 
\ifx 0\begin{comment}
  \rd{best remove it} In case the three constraints are not simultaneously feasible, whether the TCL is providing VES or not, we assume the temperature constraint has the highest priority, and then the cycling constraint, and then the energy constraint.
\end{comment} \fi
\section{Aggregate Quantities and Assumptions} \label{sec:aggQuans}
Section~\ref{sec:invTCL} was devoted to the individual TCL; we now define variables for a collection of $N$ TCLs that are needed to pose the problem precisely. The maximum possible electrical demand of the collection of $N$ TCLs is denoted by $\Ptotmax$ and the demand at $k$ is denoted by $P_k$:
\begin{align}\label{eq:Ptotmax-def}
\Ptotmax & \triangleq  \sum_{j=1}^N P^j, &  P_k \triangleq \sum_{j=1}^N P^j m_k^j.
\end{align}
The quantity  corresponding to $P_k$ during baseline operation is denoted by $P_k^b$. Recall that the collection of TCLs provide VES service by varying the individual on/off status $m_k^j$ so that the deviation of demand from the baseline tracks a grid supplied reference as closely as possible, without violating any individual's QoS constraints. The grid supplied VES reference is denoted by $R_k$, which  is the desired value of the demand deviation from baseline, denoted by $Y_k$:
\begin{align}
  \label{eq:aggPowerDev}
  Y_k \triangleq P_k - P_k^b.
\end{align}
A related quantity that will be useful later is the \emph{analytical baseline demand of the aggregate}, denoted by $\hat{P}_k^b$:
\begin{align} \label{eq:aggBaseline}
  \hat{P}_k^b \triangleq \sum_{j=1}^N   \hat{P}_k^{j,b}= \sum_{j=1}^{N}\frac{\theta_{k}^a - \theta^j_{\textSet}}{\eta^j R^j_{th}}.
\end{align}
It is the analytical counterpart to $P_k^b$. %, and could be determined by simulating a population of TCLs in baseline operation.

Our development uses the following  ``fractional'' counterparts to aggregate quantities:
\begin{align}
	n^{\textOn}_k &\triangleq \frac{\sum_{j=1}^N m_k^j}{N}, \\ 
	f_k^{\textOn} &\triangleq \frac{\sum_{j=1}^{N}F^{\textOn,j}_k}{N}, \quad 
	f_k^{\textOff} \triangleq \frac{\sum_{j=1}^{N}F^{\textOff,j}_k}{N}, \\
	%n^{\textOff}_k \triangleq \frac{\NKOff}{N}, \\
	%d^{\textOn}_k &\triangleq \frac{\sum_{j=1}^{N}D^{\textOn,j}_k}{N}, \quad 
	%d^{\textOff}_k \triangleq \frac{\sum_{j=1}^{N}D^{\textOff,j}_k}{N}, \\
	\stuckOnK &\triangleq \frac{\sum_{j=1}^{N}\stuckOnKJ}{N}, \quad
	\stuckOffK \triangleq \frac{\sum_{j=1}^{N}\stuckOffKJ}{N}.	
\end{align}
The quantity $f_k^{\textOn}$ is called the \emph{fraction at time $k$ that decide to flip on at $k+1$}, and $\stuckOnK$ is called the \emph{fraction that is stuck on at $k$}, and similarly for the ``off'' fractions.
\subsection{Role of heterogeneity}\label{sec:role-of-het}
We limit ourselves to populations in which the following assumption holds, which we call \emph{quasi-heterogeneous} populations.
\begin{assumption}\label{as:het}
  \begin{align}\label{eq:quasi-het-def}
	\text{(i):} &\quad P_k =    n_k^{\textOn}\Ptotmax. \\
	\text{(ii):} &\quad \tauTCL^1=\tauTCL^2=\dots=\tauTCL^N\delequal\tauTCL.
  \end{align}
\end{assumption}
Assumption~\ref{as:het}(i) means the fraction of loads on at $k$ is equivalent to the total power consumption at that time, and Assumption~\ref{as:het}(ii) means the lock-out constraint is the same for all the loads.

\begin{figure}
	\centering
	\includegraphics[width=1\columnwidth]{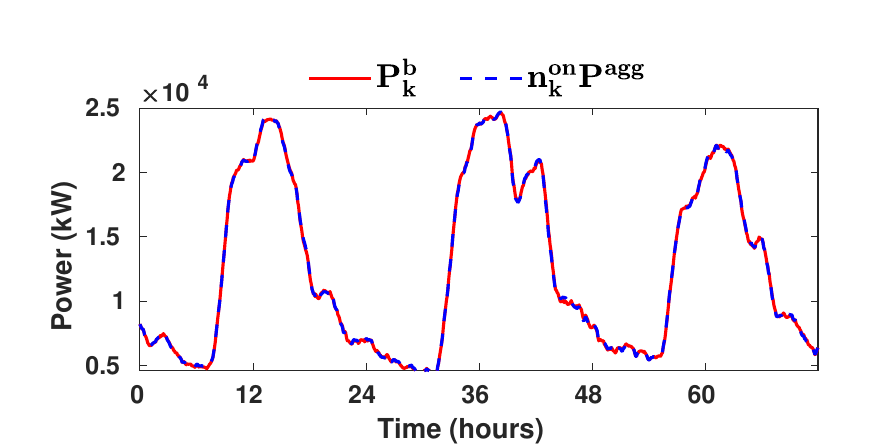}
	\caption{Comparison of $P_k$ and $n^{\textOn}_kP^{\textAgg}$,  for a heterogeneous population of TCLs with thermostat control. Simulation parameters are described in Sec.~\ref{sec:numExp}.}
	\label{fig:assumpOneVer}
\end{figure}

The assumption holds for a homogeneous population. Assumption~\ref{as:het}(i) holds approximately for a heterogeneous population if the $P^j$'s are drawn from a uniform or Gaussian distribution, or for that matter, any symmetric uni-modal distribution. Results from one numerical experiment are shown in Figure~\ref{fig:assumpOneVer}; the quantities are nearly identical in this experiment. Details of these simulations are described in Section~\ref{sec:numExp}.

The reason for introducing this assumption is that the capacity will be characterized in terms of the fraction on, $n_k^{\textOn}$, and related quantities introduced above, since they are easy to relate to the cycling constraint of individual TCLs. The Assumption~\ref{as:het}(i) allows us to translate the ensemble's power demand $P_k$ to $n_k^{\textOn}$.

 %Additionally, numerical experiments later show that the capacity characterization results of the paper hold even for heterogeneous scenarios as long as the equality in~\eqref{eq:quasi-het-def} hold approximately, such as in case of uniform or Gaussian distributed $P^j$'s. 

% We will use in our development an upper bound on the lock-out time $\tau^j$ that is denoted by $\tau^j$, i.e.,
% \begin{align}\label{eq:def-tauBA}
% \tauBA \geq \tau_j, j=1,\dots,N.
% \end{align}
\begin{figure}
	\centering
	\includegraphics[width=1\columnwidth]{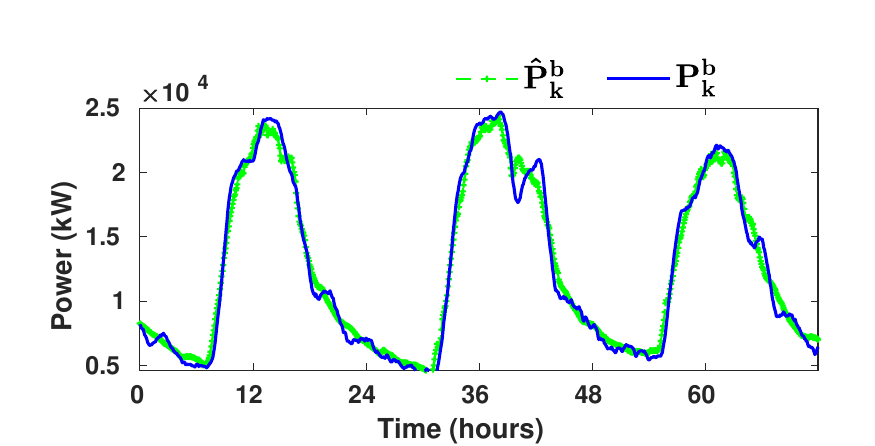}
	\caption{Comparison of $P_k^b$ and $\hat{P}_k^{b}$,  for a heterogeneous population of TCLs with thermostat control. Simulation parameters are described in Sec.~\ref{sec:numExp}.}
	\label{fig:baselineCompare}
\end{figure}

\subsection{Role of the coordination algorithm} \label{subsec:coordAlg}
For a collection of TCLs to provide VES service, the aggregate power deviation $Y_k$ of the collection has to track a reference $R_k$. A coordination algorithm is needed to perform this tracking. There are many ways to pose/design a coordination algorithm, see, for example, the references~\cite{matkoccal:2013,ChenDistributedIMA:2017,zhang2013aggregated,CoffmanVESBuildSys:2018,liushi:2016}. There are also potentially many metrics to deem a coordination algorithm well designed. While our results do not depend on a specific coordination algorithm, we do specify a requirement for a coordination algorithm to be considered well-designed. The requirement is that when the grid is not asking for VES, i.e., $R_k \equiv 0$, the coordination algorithm should mimic the baseline operation by the thermostat. This is stated formally as the following assumption.
\begin{assumption} \label{ass:anyBaseSim}
 $P_k^b = \hat{P}_k^b$.
\end{assumption}
Assumption~\eqref{ass:anyBaseSim} is used in translating QoS constraints of the individuals into a tractable constraint for the ensemble providing VES, by allowing definitions such as~\eqref{eq:aggPowerDev} to be independent of the coordination algorithm. It also allows us to use predictions of the ambient temperature to predict the baseline power, which will be useful for reference planning in Section~\ref{sec:propRefGen}. Again, the assumption needs to hold only approximately; and the reader can find numerical results justifying Assumption~\ref{ass:anyBaseSim} in Figure~\ref{fig:baselineCompare}. 

% For the second assumption, define
% \begin{align}\label{eq:deltaDK}
% 	\Delta d_k \triangleq d_k^{\textOn} - d_k^{\textOff}
% \end{align}   
% to denote the difference between the fraction of natural on and off flips due to the temperature hitting the deadband during VES operation. The quantity $\Delta d_k^b$ has the same meaning, but in baseline operation.
% \begin{assumption} \label{ass:relateDK} 
% 	%\begin{align} \nonumber
% 	$\ \sum_{k=1}^{N_t}\left|\Delta d_k\right| \leq \sum_{k=1}^{N_t}\left|\Delta d_k^b\right|.$
% 	%\end{align}
% \end{assumption}
% Assumption~\ref{ass:relateDK} says that thermostat related flips occur more frequently during baseline operation than when TCLs provide VES. When providing grid support service a TCLs will have to switch before its temperature hits the deadband, which suggests that Assumption~\ref{ass:relateDK} is likely to hold in practice. The quantity $\Delta d_k$ will appear in Section~\ref{sec:powerBounds}.

\ifx 0
\begin{coomment}
	For a homogeneous collection of TCLs, the fraction of loads on and the total power consumption are proportional. Thus in the developments to follow, ``fraction of loads on'' and ``total power consumption'' can be freely interchanged, modulo a scaling factor. 
\end{coomment}
\fi

\ifx 0        
\begin{coomment} \label{com:homApprox}
Although, $Y_k$ in~\eqref{eq:powConsump} approximates well the power consumption of a heterogeneous ensemble~\cite{ChongStatisticalTPS:1984} if there is: (i) small parameter variation and (ii) $\Expect\left[P\right] = P$ in~\eqref{eq:powConsump}. Large parametric variation (or $\tauTCL^j \neq \tauTCL$ for all $j$) can be handled by creating subgroups that are treated as approximately homogeneous (or $\tauTCL^j = \tauTCL$ for all $j$ in the group). Verification of this phenomenon is shown in Figure~\ref{fig:baselineCompare}, where a heterogeneous population of TCLs is simulated where $Y_k^{b,\text{het}} = \sum_{i=1}^{N}P^jm_k^j$ and $Y_k^b = \frac{1}{N}\big(\sum_{j=1}^{N}P^j\big)\sum_{j=1}^{N}m_k^j$; the two quantities are almost identical (see Figure~\ref{fig:baselineCompare}). 
\end{coomment}
\fi

\section{Aggregate Capacity Constraints} \label{sec:propMethod}
\textit{Aggregate capacity constraints} that will be derived in this section refers to constraints on aggregate quantities due to the temperature, cycling, and energy constraints at the individual TCL, i.e.,~\eqref{eq:invTz}-\eqref{eq:invEng}. That is, if each TCL in a collection enforces~\eqref{eq:invTz}-\eqref{eq:invEng} then the aggregate constraints will be satisfied. The contrapositive is: if the aggregate constraints are violated, there would exist at least a single TCL that violates its individual QoS constraints. Hence, violating  the aggregate constraints means that it is \emph{impossible} for every TCL to satisfy their own QoS; at least one TCL will violate its local constraints.  

\subsection{Aggregate scaled temperature} \label{sec:aggModel}
If the TCLs are homogeneous, the dynamics of the average temperature of the ensemble is the same as that of the individual, with aggregate values for the parameters in the model~\eqref{eq:dynTCL}, but that is not the case in the heterogeneous case~\cite{GuoAggregationEnB:2020}. It is still possible to develop an aggregate model that has a connection to each individual TCL's temperature constraint~\eqref{eq:invTz}, as done in~\cite{hao_aggregate:2015}, which we do next. 

Consider the aggregate demand deviation from the analytical baseline demand:
\begin{align}\label{eq:defTtildeK}
	\hat{Y}_k \triangleq P_k - \hat{P}^b_k.
\end{align}
% where $P_k$ is the total power consumption and $\hat{P}^b_k$ is the analytical baseline power consumption of the ensemble.
We call $\hat{Y}_k$ the analytical demand deviation, and it is the analytical counterpart of the actual demand deviation $Y_k$ defined in~\eqref{eq:aggPowerDev}. We have the following result.
\begin{lem}[Theorem 5 in~\cite{hao_aggregate:2015}]\label{lem:hetEnsemModel}
	For an arbitrary $\alpha > 0$, denote $\bar{a} = \exp(-T_s/\alpha)$, $\bar{b} = (1-\bar{a})\alpha$, and define
	\begin{align} \label{eq:aggEngDevDyn}
	Z_k = \bar{a}Z_{k-1} - \bar{b}\hat{Y}_{k-1}
	\end{align} 
with $Z_0=0$. If for all $j \in \{1,\dots, N\}$ and for all $k\geq0$ the constraint~\eqref{eq:invTz} is maintained with $\theta^j_0 = \theta_{\textSet}$, then  $|Z_k| \leq \tilde{C}$ for all $k$, where
	\begin{align} \label{eq:haoAggHetEng}
	\tilde{C} \triangleq \sum_{j=1}^{N}\left(1 + \left|1-\frac{R^j_{th}C^j_{th}}{\alpha}\right|\right) \frac{C^j_{th}\delta^j}{\eta^j}.
	\end{align}
      \end{lem}
      Lemma~\ref{lem:hetEnsemModel} allows us to use the bound~\eqref{eq:haoAggHetEng} as a necessary condition for each TCL to maintain the temperature constraint~\eqref{eq:invTz}. 
      \ifshowArxivAlt
      The original proof of Lemma~\ref{lem:hetEnsemModel} in~\cite{hao_aggregate:2015} is for a continuous time system with constant ambient temperature, a proof for the current setting is given in the extended arxiv version of this work~\cite{coffman2019aggregate:arxivACC}.
      \fi
      \ifshowArxiv
      The original proof of Lemma~\ref{lem:hetEnsemModel} in~\cite{hao_aggregate:2015} is for a continuous time system with constant ambient temperature. A proof for the current setting is given in the Appendix.
      \fi
      
\begin{cor} \label{cor:homEnsem}
	Let the ensemble of TCLs be homogeneous, denote the quantity,
	\begin{align} \label{eq:sumOfTemps}
		g_k = \frac{C_{th}}{\eta}\sum_{j=1}^{N}(\theta_{k}^j - \theta_{\textSet}),
	\end{align}
	and let $\alpha = C_{th}R_{th}$. Then $g_k = Z_k$ for all $k$.
\end{cor} 
That is, in the homogeneous case the quantity $Z_k$ is proportional to the temperature deviation from the setpoint, with the unit of energy (kWh thermal). While it is hard to interpret the quantity $Z_k$ in Lemma~\ref{lem:hetEnsemModel}, it is trying to capture the sum in~\eqref{eq:sumOfTemps} for a heterogeneous ensemble. We refer to the quantity $Z_k$ in the sequel as the \emph{scaled temperature deviation of the ensemble}.

      \begin{comment}\label{com:thermalEnergyConstraint}
Lemma~\ref{lem:hetEnsemModel} holds with $\hat{Y}_k$ replaced with $Y_k$. This follows immediately from Assumption~\eqref{as:het} and the definitions~\eqref{eq:aggPowerDev} and~\eqref{eq:defTtildeK}. 
      \end{comment}
      
\subsection{Fraction of TCLs stuck in on or off mode}~\label{sec:aggCycleB}
The fraction of TCL's stuck on, or off, evolves according to the following inventory model:
\begin{align} \label{eq:recurSwitch}
	\stuckOn_k = \stuckOn_{k-1} + f_{k-1}^{\textOn} - f_{k-1-\tauBA}^{\textOn}.  
\end{align}
In words, the fraction that are stuck on, $\stuckOn_{k-1}$, increases by the fraction that flip on $f^{\textOn}_{k-1}$ from $k-1$ to $k$ and decreases by the fraction that had flipped on $k-1-\tauBA$ time instants in the past. Note that Assumption~\ref{as:het} is used here: if $\tau^j$'s were distinct the equality will not hold. A similar relationship holds for the fraction stuck off: 
\begin{align} \label{eq:recurSwitch-2}
  \stuckOff_k = \stuckOff_{k-1} + \switchOff_{k-1} -\switchOff_{k-1-\tauBA}.  
\end{align}

% We define an input for stuck on (respectively, off) dynamics as the following column vector,
% \begin{align}
% 	u_{k-1}^{\textOn} \triangleq [f_{k-1}^{\textOn}, \dots,  f_{k-1-\tauBA}^{\textOn}]^T.
% \end{align}
% Eq.~\eqref{eq:recurSwitch} can now be represented as the following linear state space model:
% \begin{align} \label{eq:stuckDyn}
% \stuckOn_k  &= \stuckOn_{k-1} + Bu_{k-1}^{\textOn}, \quad \stuckOn_{0} = 0, \\ \label{eq:stuckDynEnd}
% \stuckOff_k &= \stuckOff_{k-1} + Bu_{k-1}^{\textOff}, \quad \stuckOff_{0} = 0,
% \end{align} 
% where the matrix $B$ is 
% \begin{align} 
% 	B(\tauBA) &\triangleq \begin{bmatrix}
% 	1, & \mathbf{0}_{\tauBA-2}, & -1 
% 	\end{bmatrix},
% \end{align}
% where $\mathbf{0}_{\ell}$ is a row vector of zeros of length $\ell$, and $\tauBA$ is an upper bound on $\tau_j$'s defined at the end of Section~\ref{sec:role-of-het}. 
\subsection{Fraction of TCLs on}
The fraction of TCLs on, $n_k^{\textOn}$, is a particulary important quantity since the total electrical power consumption of the ensemble at $k$ is proportional to it due to Assumption~\ref{as:het}. We now derive a dynamic model of and constraints on $n_k^{\textOn}$. This exercise does not have to be repeated for fraction off since that is completely determined by the fraction on. 
\subsubsection{Dynamics}
An  inventory equation - similar to~\eqref{eq:recurSwitch} -  couples dynamics of fraction on and fraction that flips:
\begin{align} \label{eq:switchFracRel}
n^{\textOn}_k = n^{\textOn}_{k-1} + \flipOn_{k-1} - \flipOff_{k-1}.
\end{align}
In words, the fraction of on devices at time $k$ is the fraction already on at $k-1$, plus the fraction that flipped on minus the fraction that flipped off  from time $k-1$ to $k$. 
\subsubsection{Constraints}\label{sec:powerBounds}
In the boundary case all $N$ TCLs can be on at time $k$, which means that no TCLs were previously stuck off. In the case where some TCLs were previously stuck off, an upper bound for the fraction that can be on is $n_k^{\textOn} \leq 1 - \stuckOff_{k-1}$. Similarly, since those TCLs that are stuck on at $k-1$ must be kept on at $k$, we have $n_k^{\textOn} \geq \stuckOn_{k-1}$. Thus, we have the following constraint:
\begin{align} \label{eq:refBounds}
\stuckOn_{k-1}\leq n^{\textOn}_k \leq 1 - \stuckOff_{k-1}. 
\end{align}

\ifx 0
where $B_k^{\textOn}$ refers to the power consumption of the loads at time $k$ during baseline operation. If the baseline controller is designed such that the equilibrium point specified by the baseline power given in~\eqref{eq:aggBaseline} is asymptotically stable, then asymptotically $\Delta d_k \rightarrow 0$. We take the finite time approximation that $\Delta d_k = 0$, yielding

We treat the asymptotic stability condition during baseline operation as a requirement for a coordination algorithm to be well posed. Furthermore, we accept the finite time approximation as a better solution than requiring the baseline trajectory known. The reason for this is that computing the baseline trajectory and incorporating it into reference planning is extremely non-robust. Mathematically, the effect of the finite time approximation can be analyzed for control algorithms with uniformly ultimately bounded equilibrium points and asymptotically stable equilibrium points. 
\fi

\ifx 0 The cycling constraint induces an additional constraint on the temperature:
\begin{argu} \label{argu:themEngArg}
	If a cooling TCL is stuck on at $k$ then the temperature of that TCL cannot increase at $k+1$. Contrarily, if a cooling TCL is stuck off at $k$, the temperature cannot decrease at $k+1$. 
\end{argu}
How this relates to a bound on the quantity $z_k$ is handled first for a homogeneous ensemble due to the interpretation that $z_k$ has in this case (Corollary~\ref{cor:homEnsem}). The result of this scenario informs the bound in the heterogeneous ensemble, where $z_k$ has no convenient interpretation.
\subsubsection{Homogeneous ensemble}
For the homogeneous ensemble we have,
\begin{align}
z_k = \sum_{j=1}^{N}z_k^j \leq N(1 - \stuckOnK)\bar{C} + \sum_{j \in \text{stuck on}}z_{k}^j.
\end{align}
The above is a mathematical representation of Argument~\ref{argu:themEngArg}. Now we can bound the sum as, 
\begin{align}
-N\bar{C}\stuckOnK \leq \sum_{j \in \text{stuck on}}z_{k}^j.
\end{align}
Putting the two together yields,
\begin{align} \nonumber
	\tilde{C}_k^{+,\text{hom}} = N(1 - 2\stuckOnK)\bar{C}, \ \text{and} \  \tilde{C}_k^{-,\text{hom}} = N(1 - 2\stuckOffK)\bar{C},
\end{align} 
where the bound $\tilde{C}_k^{-,\text{hom}} \leq z_k \leq \tilde{C}_k^{+,\text{hom}}$ follows. The developments for the lower bound $\tilde{C}_k^{-,\text{hom}}$ are symmetric.
\subsubsection{Heterogeneous ensemble} 
A more conservative bound for~\eqref{eq:haoAggHetEng} is 
\begin{align} \label{eq:haoAggEng}
	 |z_k| \leq \sum_{j=1}^{N}\bar{C}^j = N\bar{C}, \ \text{where} \ \bar{C} = \frac{1}{N} \sum_{j=1}^{N}\bar{C}^j.
\end{align}
%which is the bound for a homogeneous collection computed with the average parameter values. %However, when individual TCLs have cycling constraints, both bounds~\eqref{eq:haoAggHetEng} and~\eqref{eq:haoAggEng} are not tight when the cycling constraint~\eqref{eq:invCycle} is ``active,'' i.e., some TCLS in the population are stuck on or off. %In what follows, we adopt the bound~\eqref{eq:haoInvEng} as the thermal energy bound for a TCL that is \textit{not} stuck on or off.
%The individual cycling constraint~\eqref{eq:invCycle} is argued to effect the bound for the thermal energy deviation of a TCL stuck on or off. We then proceed to show that this translates to a bound on the aggregate thermal energy deviation. The following argument assumes that when a TCL is on the temperature decreases and when a TCL is off the temperature increases.
%If a TCL is not stuck on/off then the bound on its thermal energy deviation satisfies,
%\begin{align}
%|z_{k}^j| \leq \bar{C} = \frac{C_{th} \Delta}{2\eta}.
%\end{align}
%Since this bound is on the absolute value of $z_{k}^j$ it implies that a TCL, at any point in time, could either increase or decrease its thermal energy deviation. However, this is only true if a TCL is not stuck on or off. 
Now, we split the bound~\eqref{eq:haoAggEng} into the upper and lower bounds respectively, 
\begin{align} \label{eq:socHetUpBound}
C_k^+ &= N(1 - \stuckOnK)\bar{C} + N\stuckOnK\bar{C}, \\ \label{eq:socHetLowBound}
C_k^- &= -N(1 - \stuckOffK)\bar{C} -N\stuckOffK\bar{C}. 
\end{align}
We then elect the following quantities,
\begin{align} \nonumber
\tilde{C}_k^{+} = N(1 - \stuckOnK)\bar{C}, \ \text{and} \ 
\tilde{C}_k^- = -N(1 - \stuckOffK)\bar{C}, 
\end{align}
to bound $z_k$ as,
\begin{align} \label{eq:engBound}
	\tilde{C}_k^- \leq z_k \leq \tilde{C}_k^+.
\end{align}
Through construction we have that $C_k^- \leq \tilde{C}_k^- \leq z_k \leq \tilde{C}_k^+ \leq C_k^+$ and hence $z_k$ will also satisfy~\eqref{eq:haoAggHetEng}. 
%Note, that while the bound~\eqref{eq:engBound} is true, we have not directly used Argument~\ref{argu:themEngArg} to develop this bound as we did for a homogeneous ensemble. Although, the ensemble being heterogeneous should not effect the premise of Argument~\ref{argu:themEngArg}. Hence, the construction of a bound on $z_k$ that closely resembles the bound for a homogeneous ensemble where Argument~\ref{argu:themEngArg} is used. \pb{Austin, these types of statements usually appear as red flags to the reviewers. Then they end up having the opposite effect of what you hoped for.}
\fi
\subsection{Aggregate capacity characterization}
%We collect the aggregate scaled temperature dynamics~\eqref{eq:aggEngDevDyn} with the aggregate power deviation constraint~\eqref{eq:refBounds} and thermal energy deviation constraint~\eqref{eq:haoAggHetEng} to form an aggregate constraint set on the demand deviation $Y_k$.
From Assumption~\eqref{as:het}, the demand deviation $Y_k$ is related to the fraction of loads on $n_k^{\textOn}$, which is also related to fraction stuck on/off and fraction flipped through~\eqref{eq:recurSwitch}-\eqref{eq:recurSwitch-2} and~\eqref{eq:switchFracRel}, respectively. Each of these signals have constraints and some have dynamics, which were derived in previous sections. These are now collected to describe all the constraints on the signal $Y_k$ in order to satisfy TCLs' local QoS. We first ``lift'' the signal $Y_k$, for $ k=t+1,\dots,t+H_p$ over a planning horizon $H_p$ to a decision vector $\psi_t^{t+H_p-1}$ that is defined as
\begin{align}\label{eq:psi-def}
\psi_t^{t+H_p-1} \triangleq \Big[&\{Z_k\}_{t+1}^{t+H_p}, \ \{Y_k\}_{t+1}^{t+H_p},\ \{\flipOn_k\}_t^{t+H_p-1}, \dots \nonumber \\ &\{\flipOff_k\}_{t}^{t+H_p-1},\ \{\stuckOn_k\}_{t+1}^{t+H_p}, \ \{\stuckOff_k\}_{t+1}^{t+H_p} \Big]. 
\end{align}
The capacity of the ensemble, the admissible $\{Y_k\}_{k=t+1}^{t+H_p}$ is obtained in terms of the expanded signal $\psi_t$. Specifically, given a baseline demand $\hat{P}_k^b$ over the same horizon, \emph{the capacity of the collection is the set of $\psi_t^{t+H_p-1}$'s that lie in the set $\Omega_t^{t+H_p-1}$, where} 
\begin{align} \label{eq:fixInitCond} 
\Omega_t^{t+H_p-1} \triangleq\bigg\{ &\psi_{t}^{t+H_p-1} \bigg\vert \   
                                    Z_t=0, \ \stuckOff_t = 0, \ \stuckOn_t = 0, \\ \nonumber
  & Y_t = 0, \ \text{for all} \quad k \in \{t,\dots,t+H_p-1\}, \\                                  \label{eq:battModel-in-Omega}
  & Z_{k+1} = \bar{a}Z_{k} - \bar{b}Y_{k}, \quad \left|Z_{k+1}\right| \leq \tilde{C}, \\
& n_k^{\textOn} = \frac{1}{\Ptotmax}(Y_k + \hat{P}^b_k), \label{eq:nkon-in-Omega}\\ \label{eq:nkon-bound-in-Omega}
&\stuckOn_{k} \leq n_{k+1}^{\textOn}  \leq 1 - \stuckOff_{k}, \\ \label{eq:skon-dyn-in-Omega}
&\stuckOn_{k+1} = \stuckOn_{k} + \flipOn_{k} - \flipOn_{k-\tauBA}, \\ \label{eq:skoff-dyn-in-Omega}
&\stuckOff_{k+1} = \stuckOff_{k} +\flipOff_{k}-\flipOff_{k-\tauBA}, \\ \label{eq:nkon-dyn-in-Omega}
                         &n_{k+1}^{\textOn} = n^{\textOn}_{k} + \flipOn_{k}-\flipOff_{k},\\ \label{eq:ineq-cons-in-Omega}
                         &  n_k^{\textOn},\stuckOn_k, \stuckOff_k,\flipOn_k, \flipOff_k \in [0,1], \\
  & \sum_{k=t}^{t+H_p} Y_k = 0 \bigg\}. \label{eq:aggZeroEng}
\end{align}
Recall that the constants $\tilde{C},\Ptotmax$ are defined in~\eqref{eq:haoAggHetEng}, \eqref{eq:Ptotmax-def}, and the signal $\hat{P}^b_k$ in~\eqref{eq:aggBaseline}. Eq.~\eqref{eq:nkon-in-Omega} uses Assumptions~\ref{as:het} and~\ref{ass:anyBaseSim}. The last constraint~\eqref{eq:aggZeroEng} acts as a necessary condition for the QoS constraint~\eqref{eq:invEng} for any collection of positive numbers $\{\tilde{E}^j\}$ and any  $H_p$ that satisfiew $H_p \leq H_b^j$. %Recall that $H_b$ can be on the order of a month. In contrast,  we expect the planning horizon $H_p$ will be much shorter since prediction of the grid's requirement, $R_k^{BA}$, is unlikely to be available for long time durations.

The following result is useful when the constraint set $\Omega_t^{t+H_p-1}$ is used to perform reference planning.

\begin{lem}\label{lem:optProb}
	The  set $\Omega_t^{t+H_p-1}$ is convex for every $t$ and $H_p\geq 1$.	Suppose that for a given $\tauBA$ and $H_p$ for all $t$, the following signal
	\begin{align}
	\bar{\theta}^a_k \triangleq \rho\theta^a_k, \quad \text{with} \quad \rho \triangleq \sum_{j=1}^{N}\frac{1}{\eta^j R^j_{th}}\bigg(\sum_{j=1}^{N}P^j\bigg)^{-1}	
	\end{align}
	satisfies
	\begin{align}
	\Theta^-_{k}(\tauBA) + \Gamma \leq \bar{\theta}^a_{k+1} \leq 1 - \Theta^+_{k}(\tauBA) + \Gamma,
	\end{align}
	for $k \in \{t, \dots, t+H_p - 1\}$,	where
	\begin{align}
	\Theta^-_{k}(\tauBA) &= \sum_{s = k - \tauBA+1}^{k}\max\{\bar{\theta}^a_{s} -\bar{\theta}^a_{s-1},0 \} \\
	\Theta^+_{k}(\tauBA) &= \sum_{s = k - \tauBA+1}^{k}\max\{\bar{\theta}^a_{s-1} -\bar{\theta}^a_{s},0 \}, \quad \text{and} \\
	\Gamma &= \sum_{j=1}^{N}\frac{\theta_{\textSet}^j}{\eta^j R^j_{th}}\bigg(\sum_{j=1}^{N}P^j\bigg)^{-1}.
	\end{align}
	Then the set $\Omega_t^{t+H_p-1}$ is non-empty for every $t$ and $H_p\geq 1$.
\end{lem}
\ifshowArxivAlt
\begin{proof}
	See extended arxiv version~\cite{coffman2019aggregate:arxivACC}.
\end{proof}
\fi
\ifshowArxiv
\begin{proof}
	See appendix.
\end{proof}
\fi

The condition on the ambient temperature in Lemma~\ref{lem:optProb} is technical: we have never run into a numerical example (with time varying $\theta^a_k$) where the result of the Lemma does not hold. An example of an ambient temperature trajectory that satisfies this assumption is a constant trajectory. We emphasize that none of the results in this section require the ambient temperature to be constant. 

\ifx 0
\begin{coomment}
The fraction of on devices $n^{\textOn}_k$ is related to the power deviation of the collection by $y_k = n^{\textOn}_kP_{\textAgg} - \bar{P}$. The point here is that the developed equations are simply linear transformations of the aggregate power deviation. Furthermore, they describe the fraction of devices that \textit{would} be either stuck on or off, given the collection followed a certain power deviation trajectory. 
\end{coomment}
\fi
\ifx 0
\begin{coomment}
The relationship between the fraction of devices on and the fraction of on and off switches~\eqref{eq:switchFracRel} is a quantity independent from the coordination algorithm. Meaning, regardless of how the population of TCLs is controlled, the fraction on can be thought of as a dynamic discrete time system with the fractional switching differential as input.
\end{coomment}
\fi

\section{Reference \optProbName} \label{sec:propRefGen}
\textit{Reference \optProbName} utilizes the aggregate capacity set from Section~\ref{sec:propMethod} to plan a reference power deviation trajectory for an ensemble of TCLs to track so that the planned reference is within the TCL's capacity. At time $t$, this is done by projecting the BA's \emph{total desired demand deviation}, $\{R_k^{BA}\}_{k=t}^{t+H_p-1}$, onto the aggregate capacity set $\Omega_{t}^{t+H_p-1}$ to obtain the optimal $\psi^*$. We need the following definition:
\begin{align} \nonumber
  (\psi^{\textBA})_{t}^{t+H_p-1} & \triangleq \Big[\{0\}_{t+1}^{t+H_p}, \{ R_k^{\textBA} \}_{t+1}^{t+H_p}, \{0\}_t^{t+H_p-1}, \nonumber \\
   & \quad \{0\}_t^{t+H_p-1}, \{0\}_{t+1}^{t+H_p}, \{0\}_{t+1}^{t+H_p} \Big].
\end{align}
The \textit{reference \optProbName} problem can be cast as the following convex optimization problem,
\begin{align} \label{eq:optProb}
  \begin{split}
	\psi^* = \arg \min_{\psi} & \ J(\psi) = \|\psi^{\textBA}-\psi\|^2_{\Xi} \\ 
	&\text{s.t.}\quad \psi \in \Omega    
  \end{split}
\end{align}
where sub/super-scripts are omitted from $\psi,\psi^*$ to reduce clutter, $\Xi$ is a symmetric positive definite (s.p.d.) weighting matrix of appropriate dimension, and for $x \in \R^n$,  $\|x\|^2_Q \eqdef x^TQx$ for a s.p.d. $n \times n$ matrix $Q$.

The component $\{Y_k^*\}$ of $\psi^*$ -- see the definition~\eqref{eq:psi-def} --  is denoted by $R_k^*$ in the sequel: it is the ``largest'' power deviation reference, aligned with the BA's needs, that the TCLs can track without any TCL having to violate its QoS constraints. 

The objective function $J(\psi)$ is strictly convex since $\Xi$ is a s.p.d matrix. Combining this with Lemma~\ref{lem:optProb}, we have that a solution to the reference planning problem always exists and is unique. In other words, for any $\psi^{BA}$ there will always exist a unique reference signal that a collection of TCLs are ideally suited to track. 

\subsubsection{Information requirement} In order for a BA to solve the reference planning problem~\eqref{eq:optProb}, it needs to know: (i) the parameters $P^{\textAgg}, \tauBA$, $\tilde{C}$, $\bar{a}$, and $\bar{b}$ (ii) the initial conditions $n_t^{\textOn},\flipOn_{t-1},\dots, \flipOn_{t-\tauBA}, \flipOff_{t-1},\dots, \flipOff_{t-\tauBA}$, and (iii) forecasts of the signals $\theta_k^a, R_k^{\textBA}$ over the planning horizon $H_p$.  The ambient temperature forecast can be obtained from weather services and the forecast of $R_k^{BA}$ can be obtained from a prediction of the net load~\cite{bar:2019}. In the numerical simulations conducted later, we set the initial condition $n_t^{\textOn}$ to $\hat{P}^b_k/P^{\textAgg}$ (which corresponds to $Y_t = 0$ as prescribed in $\Omega_t^{t + H_p-1}$). The initial fraction of loads stuck on/off and the initial scaled temperature deviation are assumed to be zero, as specified in~\eqref{eq:fixInitCond}. Alternatively, the BA could obtain these quantities through measurements from the population of TCLs. 

\ifshowArxiv
\subsection{Alternative Method for Reference Planning} \label{sec:altRefGen}
To compare with past literature we define a constraint set based on the constraints developed in~\cite{hao_aggregate:2015} and the scaled aggregate temperature deviation model~\eqref{eq:aggEngDevDyn} for projection of $R_k^{\textBA}$. The disadvantage with this constraint set is that the aggregate power and scaled temperature deviation bounds developed in~\cite{hao_aggregate:2015} do not account for the individual cycling~\eqref{eq:invCycle} or energy~\eqref{eq:invEng} constraint. This alternative reference \optProbName \ problem is posed as
\begin{align} \label{prob:heHao}
&\min_{\{Y_k\},\{Z_k\}} \ \sum_{k=t}^{t+H_p-1}(R_k^{\textBA} - Y_k)^2\xi + \sum_{k=t+1}^{t+H_p}Z_k^2\\ \nonumber
&\qquad \quad \text{s.t.} \quad \forall \ k \in \{t,...,t+H_p-1\} \\ 
&\qquad \quad Z_{k+1} = \bar{a}Z_k - \bar{b}Y_k, \quad Z_t = 0, \\ \label{eq:noCycBounds}
&\qquad \quad |Z_{k+1}| \leq \bar{C},  \quad -\hat{P}^b_k \leq Y_k \leq P_{\textAgg} - \hat{P}^b_k,
\end{align} 
where $\xi$ is a constant that specifies the relative importance of goals in the objective. If compared to the bounds developed in Section~\ref{sec:propMethod}, the bounds for $Z_k$ and $Y_k$ in~\eqref{eq:noCycBounds} assume that no TCLs have lock out constraints. 
\fi

\ifx 0
\begin{coomment}
	From the proof of Lemma~\ref{lem:optProb} the vector $\psi = 0$ is in the total constraint set. So setting the ``projection'' elements, excluding $r_k^{\textBA}$, of  $\psi^{\textBA}$ to zero is equivalent to desiring the other decision variables to ``stay'' within the set. Practically, the solution obtained will best track $r_k^{\textBA}$ while also minimizing the fraction of on/off switches, the fraction stuck on/off, and the aggregate thermal energy. The relative magnitude of the diagonal elements in $\Xi$ specify the level of preference for each of these goals; the decision variable that corresponds to the largest value in $\Xi$ will take precedence in minimizing its distance to the corresponding element in $\psi^{\textBA}$.
\end{coomment}
\fi

\section{Numerical Experiments} \label{sec:numExp}
We survey here numerical experiments conducted with our proposed reference \optProbName \ method, and compare the results with those from the alternative method that is representative of the prior art. The simulated TCLs are residential air conditioner units (ACs). The alternative method is designed to satisfy indoor temperature constraint but  \emph{does not} account for cycling constraints of the TCLs. 
\ifshowArxivAlt
For a full description of the alternative method see~\cite{hao_aggregate:2015} and for a description of how to use it to plan reference trajectories see~\cite{coffman2019aggregate:arxivACC}.
\fi
\ifshowArxiv
For a full description of the alternative method see~\cite{hao_aggregate:2015}, where a description of how to use it to plan reference trajectories was given in Section~\ref{sec:altRefGen}.
\fi

Both the proposed method and the alternate method return a reference trajectory for the ensemble. Both methods involve the solution of a convex optimization problem, which is performed using CVX~\cite{cvx}.

We also present closed loop simulations. The purpose is to illustrate that the trajectory computed with the proposed method is \emph{within the capacity of the TCLs}, meaning that the TCLs can collectively track it without any TCL having to violate its QoS constraints. In contrast, we will show that the reference from the alternate method is beyond the capacity of the ensemble; some of the TCLs will have to violate their local QoS constraints in order to collectively track the reference. Alternately, if local QoS is enforced by a local controller, the ensemble will not be able to track the reference. This is demonstrated by performing closed loop simulation with a centralized controller to coordinate the TCLs to track the planned reference signal. The centralized coordinator is a priority stack controller: It is a modified version of the one presented in~\cite{hao_aggregate:2015}. While the original one described in~\cite{hao_aggregate:2015} enforces the temperature QoS of each TCL, i.e.,~\eqref{eq:invTz}, the modified coordinator presented here also enforces each TCL's cycling QoS~\eqref{eq:invCycle}, but not the energy QoS~\eqref{eq:invEng}.

The closed loop simulation are performed for three reference tracking scenarios: (t-i) reference computed from the proposed method, (t-ii) reference computed from the alternative method, and (t-iii) reference from the alternative method, but the coordinator \emph{does not} enforce the cycling constraint of the TCLs. \emph{We find that only in scenario (t-i) will the ensemble of TCLs be able to track the planned reference while each individual maintaining all three of its QoS constraints.} Details are described next.

%\begin{coomment}
% We used a priority stack controller in these experiments due to its simplicity. Coordination algorithms that depend on multitudes of tuning parameters could obscure the results if these hyper parameters are not tuned correctly.
%\end{coomment}
\def\arraystretch{1.4}
\begin{table}[t]
	\centering
	\caption{Simulation Parameters}
	\label{tab:BP}
	\begin{tabular}{|| l |c| c|| l |c| c||}
		Par. & Unit & value & Par. & Unit & value \\
		N & thousand & 60 & $\eta^j$ & N/A & $2.5$ \\
		$\bar{C}$ & MWh & 50 & $\theta^a_k$ & $^{\circ}$C & time var. \\
		$\tauBA$ & Mins. & 20 & $\theta_{\textSet}^j$ & $^{\circ}$C & $U[21,21.4]$\\
		$\tauTCL$ & Mins. & 10 & $\delta^j$ & $^{\circ}$C & $U[0.75,1]$\\
		$R_{th}^j$ & $^{\circ}$C$/$kW & $U[2,2.4]$ & $T_s$ & Mins. & 2\\
		$C_{th}^j$ & kWh$/^{\circ}$C & $U[2,2.4]$ & $P^j$ & kW & $U[5.6,7]$\\
		$\tilde{E}^j$ & kWh & 6.4 & $P^{\textAgg}$ & MW & 134.4\\
		%$\eta$ & $N/A$ & 2.5 \\
		%$T_a$ & $^{\circ}C$ & 30 \\
		%$\theta_{set}$ & $^{\circ}C$ & 21 \\
		%$\Delta$ & $^{\circ}C$ & 2 \\
		%$T_s$ & Mins. & 2 \\
	\end{tabular}
	
	$^*U[a,b]$ represents uniform distribution on $[a,b]$.
\end{table}
  
\subsection{Reference Planning}  
\begin{figure}
	\centering
	\includegraphics[width=1\columnwidth]{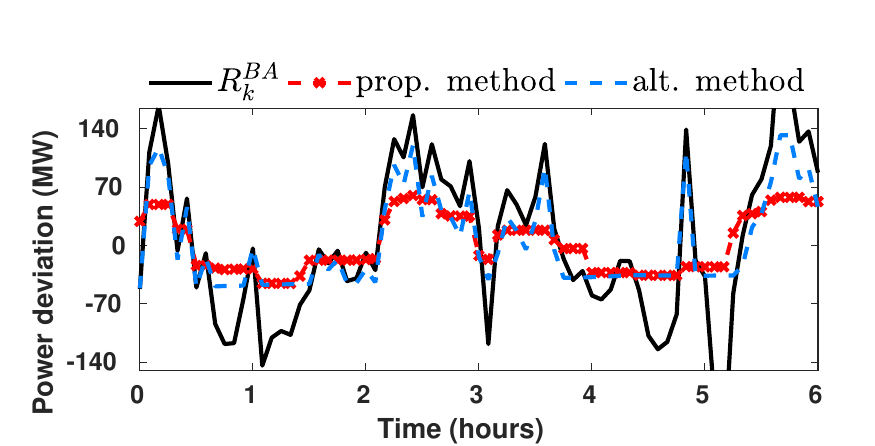}
	\caption{BA signal ($R_k^{BA}$) and the reference trajectories ($R^*_k$) for a collection of $60,000$ TCLs.}
	\label{fig:simData}	
\end{figure}

For both reference \optProbName \ methods the BA supplied reference, $R_k^{\textBA}$, is obtained from BPA, a Balancing Authority in the Pacific Northwest of the United States, and is shown in Figure~\ref{fig:simData}. A heterogeneous ensemble of loads is considered. The parameters for the loads are based on the values provided in~\cite{mathieu_revenue} and these values are shown in Table~\ref{tab:BP}, along with other simulation parameters. The ambient air temperature is time varying; it is obtained from weatherunderground.com for a summer day in Gainesville, Fl. Each TCL experiences the same ambient temperature.

Figure~\ref{fig:simData} shows the reference signals planned by the two methods, the proposed method and the alternate one. We plan both references for one day, but only show a portion of the results in Figure~\ref{fig:simData} for clarity; tracking results in the next section are shown for the full horizon. The reference signal planned with the proposed method is noticeably less aggressive than the reference signal planned with the alternative method. That is, when cycling constraints are not taken into account higher ramp rates are asked from the collection of TCLs to get closer to the BA's requirement. As we will see shortly, this leads to either poor reference tracking, violation of individual TCL's QoS, or both.

\subsection{Closed Loop Reference Tracking}
\begin{figure} [h]
	\centering
	\includegraphics[width=1\columnwidth]{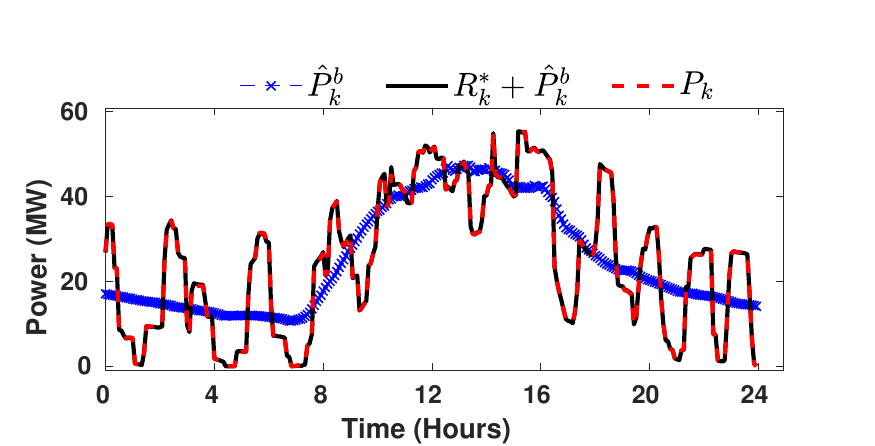}
	\includegraphics[width=1\columnwidth]{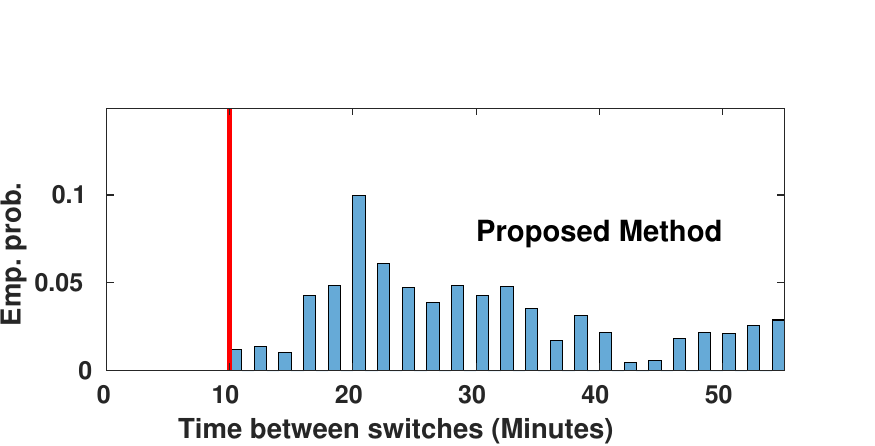}
	\caption{Closed loop results in scenario t-i: reference planned from the proposed method. (Top): reference tracking results, (Bottom): individual TCL cycling QoS results. The vertical red line indicates \tauTCL.}
	\label{fig:powerTrackACs}
\end{figure}

\paragraph{Scenario t-i} The closed loop output $P_k$ is shown in Figure~\ref{fig:powerTrackACs} along with the reference planned with the proposed method. The collection of AC units are able to track the planned reference signal with minimal tracking error (see Table~\ref{tab:resultsRefTrack}). The individual cycling QoS results are shown in Figure~\ref{fig:powerTrackACs} (bottom). Every AC satisfies its cycling QoS: No units cycle faster than $\tauTCL = 10$ minutes and the majority of the cycling times concentrate near $\tauBA = 20$ minutes.

\paragraph{Scenario t-ii}
The closed loop output $Y_k$ is shown in Figure~\ref{fig:powerTrackACs}, along with the reference (planned by the alternative method that does incorporate cycling constraints). Since this reference is beyond the capacity of the TCLs, and the coordinator enforces cycling QoS at the individuals, the collection of AC units track the planned reference poorly. For comparison, the reference tracking error reported in Table~\ref{tab:resultsRefTrack} is \emph{two orders of magnitude} higher than the error with our proposed method. This illustrates the need for TCL's cycling constraints to be incorporated in reference planning. 

\begin{figure} [h]
	\centering
	\includegraphics[width=1\columnwidth]{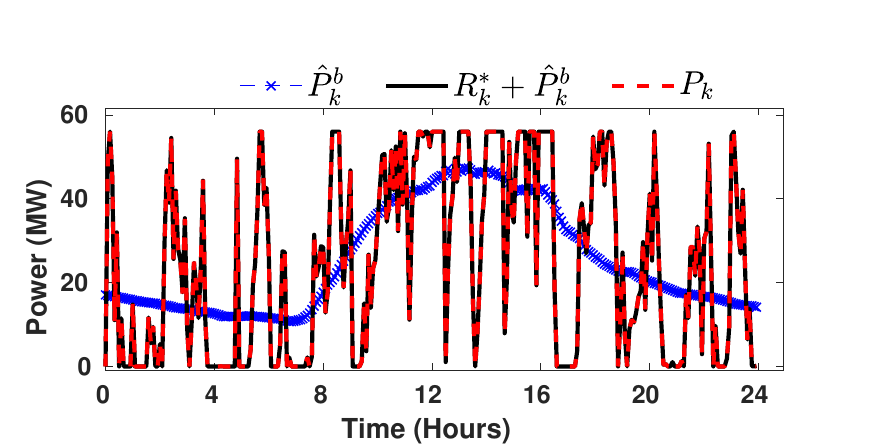}
	\includegraphics[width=1\columnwidth]{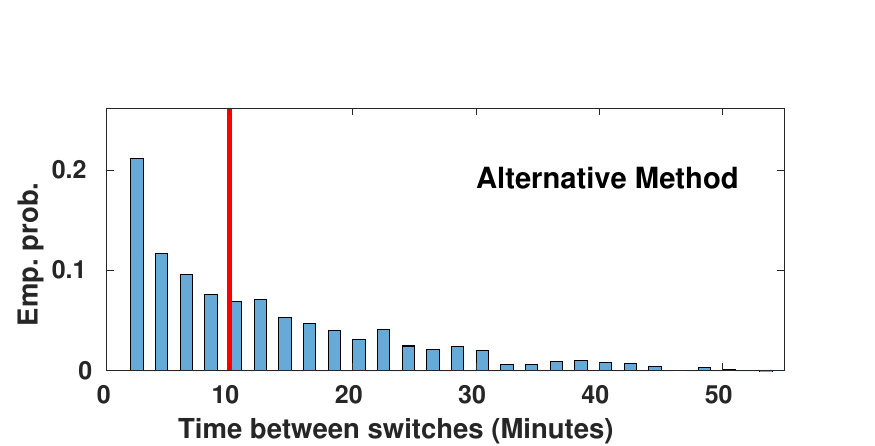}
	\caption{Closed loop results in scenario t-iii: reference planned from the alternative method and the coordinator TCLs \emph{does not} enforce TCL's cycling constraint. (Top): reference tracking, (Bottom): individual TCL cycling QoS. The vertical red line indicates \tauTCL.}
	\label{fig:powerTrackACs_noCycle_double}
\end{figure}

%Another consequence of the reference from the alternative method neglecting the capacity is that this actually \emph{prevents} the priority stack controller from enforcing the cycling QoS, Figure~\ref{fig:powerTrackACs_noCycle} (bottom). The reference signal is requiring TCLs to switch on or off too close to the deadband, so that when a TCL switches to enforce~\eqref{eq:invTz} it will have switched in a time less than \tauTCL \  from its previous switch. %Interestingly, this suggests that when a reference signal exceeds the cycling capacity that just to enforce the cycling QoS an a-causal controller is required, e.g. a predictive control scheme.
\paragraph{Scenario t-iii} Results are shown in Figure~\ref{fig:powerTrackACs_noCycle_double}: good reference tracking at the cost of excessive cycling. Roughly 20 $\%$ of the total switches occurring 2 minutes apart (the sampling time). %From experience this result is consistent across sampling times; the constraints in the alternative method assume the ability of the TCLs to switch at the sampling time. 

\begin{table}
	\centering
	\caption{Reference Tracking Errors}
	\setlength{\arrayrulewidth}{0.03cm}
	\label{tab:resultsRefTrack}
	\begin{tabular}{l | c }
		\hline
		Reference \optProbName \ method & \multicolumn{1}{c}{Tracking Error} \\ 
		\hline
		Proposed method (Figure~\ref{fig:powerTrackACs})& 0.06 $\%$ \\
		Alternative method (Figure~\ref{fig:powerTrackACs_noCycle})& 24 $\%$ \\
		\hline
	\end{tabular}
\end{table} 

\begin{figure} [h]
	\centering
	\includegraphics[width=1\columnwidth]{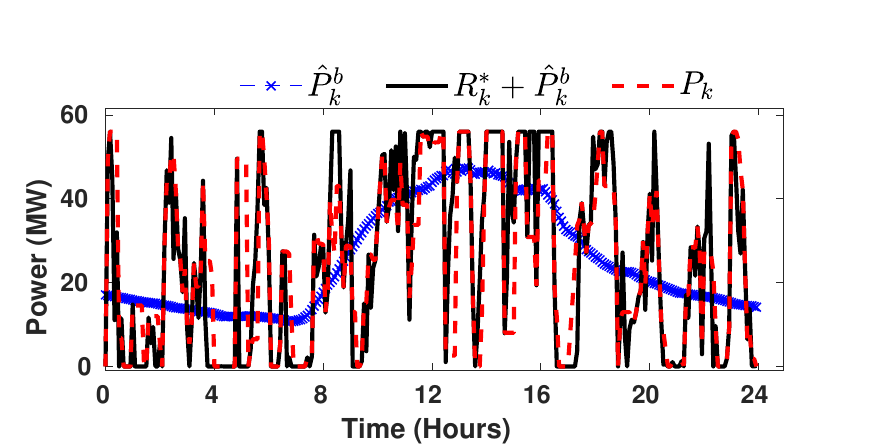}
	\includegraphics[width=1\columnwidth]{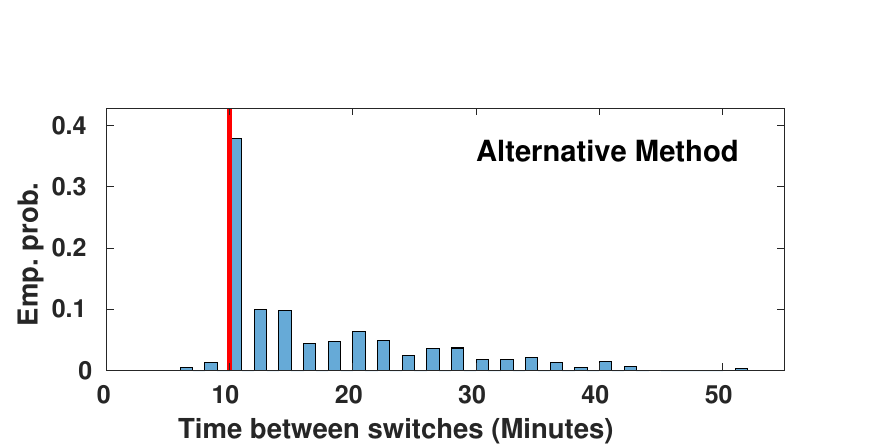}
	\caption{Closed loop results in scenario t-ii: reference planned from the alternative method. (Top): reference tracking, (Bottom): individual TCL cycling QoS. The vertical red line indicates \tauTCL.}
	\label{fig:powerTrackACs_noCycle}
\end{figure}

\ifx 0 
 
\subsection{Parametric Study}
In this study $r_k^{BA}$ is the same as the method comparison scenario and the parameters are as shown in Table~\ref{tab:BP}, except for $\tauTCL$ and $\tauBA$ which are varied over a range. To proceed with the parametric study over $\tauTCL$ and $\tauBA$, we define the metrics, $s^{\tau}$ and $d^{\tau}$. The first is, $s^{\tau}$:
\begin{align}
	 s^{\tau} \triangleq \frac{1}{N}\sum_{j=1}^{N}s^j, \ \text{where} \ s^j \triangleq \sum_{k=1}^{N_t}\left|m_k^j-m_{k-1}^j\right|,
\end{align}
which counts the total number of TCL switches normalized by the number of TCLs. The desired value of $s^{\tau}$ is small, as a smaller number of total switches is preferred, but some amount of switching is required for providing VES and maintaining thermal comfort~\eqref{eq:invTz}. The second metric is,
\begin{align}
	d^{\tau} &\triangleq \sum_{k=1}^{N_t}H_k,
\end{align}
where the variable $H_k$ is defined as,
\begin{align} \nonumber
	H_k \triangleq \begin{cases}
	1, \quad (1-\gamma_{k-1}^{\textOff} + \Delta d_{k-1}) < \frac{r_k + \bar{P}}{P_{\textAgg}} &\text{and} \quad (r_k - y_{k-1}) > 0. \\
	1, \quad (\gamma_{k-1}^{\textOn} - \Delta d_{k-1}) > \frac{r_k + \bar{P}}{P_{\textAgg}}  &\text{and} \quad (y_{k-1} - r_k) > 0. \\
	0, \quad \text{otherwise}.
	\end{cases}
\end{align}
$H_k$ is 1 if the current fraction of TCLs on, $(y_{k-1}+\bar{P})/P_{\textAgg}$, is required to increase or decrease to a value, $(r_{k}+\bar{P})/P_{\textAgg}$, that is beyond the derived limits in~\eqref{eq:refBoundsOne} and $0$ otherwise. Thus this metric counts the total number of times that the aggregate power capacity is exceeded. The desired value of $d^{\tau}$ is zero.

The two metrics are computed as follows: (i) use our proposed method and a given $\tauBA$ value to plan a reference signal, $\{r_k\}_{k=0}^{N_t-1}$(ii) use the priority stack controller to coordinate an ensemble of simulated TCLs with cycling metric $\tauTCL$ to track $r_k$ and (iii) after the simulation collect the relevant data for computation of $s^{\tau}$ and $d^{\tau}$. 

The values of the two metrics for various values of $\tauTCL$ and $\tauBA$ are shown in Figure~\ref{fig:tauMetrics}, and the results indicate that when $\tauBA$ is increased both $s^{\tau}$ and $d^{\tau}$ are decreased. In other words, the use of $\tauBA > \tauTCL$ is working as expected. 

From the results of the parametric study it is also clear that $\tauBA > \tauTCL$ is necessary if $\Delta d_k$ is to be removed from the power bound~\eqref{eq:refBoundsOne}. For instance, consider a scenario for which TCLs implement $\tauTCL = 15$ (30 mins.). If the reference is designed with $\tauBA = 15$ there are points in time where the capacity is exceeded (as indicated in Figure~\ref{fig:tauMetrics}, bottom). Although, if the reference were designed with $\tauBA = 40$ the capacity is never exceeded for $\tauTCL = 15$. 

The choice $\tauBA > \tauTCL$ was also included to reduce the total number of switches for a TCL. The success of this is documented in the top of Figure~\ref{fig:tauMetrics}, as when $\tauBA$ is increased $s^{\tau}$ is decreased. As another example, if the desired opt-out time is $\tauTCL = 5$, designing a reference with $\tauBA = 15$, instead of $\tauBA = 5$, decreases $s^{\tau}$ from $22$ to $10$. Roughly, this means that the average TCL engages in half the amount of mode state switches over the same given time horizon.

\begin{figure}
	\centering
	\includegraphics[width = 1\columnwidth]{tauVsSmetric.pdf}
	\includegraphics[width = 1\columnwidth]{tauVsDmetric.pdf}
	\caption{The metrics $s^{\tau}$ (top) and $d^{\tau}$ (bottom) against $\tauTCL$, for various values of $\tauBA$.}
	\label{fig:tauMetrics}
\end{figure}
\fi
\section{Summary and conclusion} 
The aggregate capacity characterization proposed here takes into account temperature, cycling, and energy use constraints at each individual TCL. The characterization is in the form of a set of constraints on aggregate quantities. These constraints can be thought of as necessary conditions: if the aggregate state variables for the collection violates these constraints, at least one TCL will have to violate its QoS constraints. Numerical experiments show that the cost of ignoring some of the QoS constraints in the capacity characterization - a feature of prior work - is high: the alternative characterization that does not include cycling constraints leads to tracking errors two orders of magnitude higher than the proposed one. 

The information needed to set up the reference planning problem include parameters representing an average TCL such as its COP, allowable temperature bounds, etc. Numerical experiments indicate the results are quite robust to the quasi-homogeneity assumption. It remains to be explored how heterogeneous a collection has to be before the characterization provided is no longer useful. 

The reference planning problem we examined here is a short-term planning problem: its problem data includes prediction of mismatch between demand and supply (in MW). An open problem is capacity characterization of TCLs for long-term planning. An investigation for flexible loads that do not have cycling constraints is provided in~\cite{CoffmanSpectral_ACC:2020}.

%\section*{Acknowledgement}
%

\ifx 0
\section{Acknowledgments}
	The research reported here has been partially supported by the NSF through award 1646229 (CPS-ECCS). A. Coffman is partially supported by University of Florida' Graduate School Preeminence Award.
 \fi
 %\balance
 \ifshowArxivAlt
\bibliographystyle{IEEEtran}
\bibliography{\DiCEbibPATH/grid,\JBbibPATH/Jonathan,\DiCEbibPATH/systemid,\DiCEbibPATH/Barooah,\DiCEbibPATH/optimization,\DiCEbibPATH/basics}
\fi

\ifshowArxiv

\fi

\ifshowArxiv
\section*{Appendix}
\subsection{ Proof of Lemma~\ref{lem:hetEnsemModel}}
	The proof roughly follows the one found in~\cite{hao_aggregate:2015}, with slight modification to handle time varying weather and the discrete time dynamics. By construction, the discrete time dynamics~\eqref{eq:aggEngDevDyn} is the discrete time equivalent of the ode
	\begin{align}
		\dot{z}(t) = -\alpha z(t) - \tilde{y}(t)
	\end{align}
        with zero-order-hold,	where $Z_k = Z(t_k)$ and $\tilde{Y}_k = \tilde{Y}(t_k)$. A similar observation is true for the recursion~\eqref{eq:dynTCL}. That is, the discrete time dynamics~\eqref{eq:dynTCL} is the discrete-time equivalent of the ode
	\begin{align}
		\dot{\theta}(t) = \frac{1}{RC}\left(\theta^a(t) - \theta(t)\right) + \frac{\eta}{C}P(t), 
	\end{align}
with zero-order hold,	where $\theta_k = \theta(t_k)$, $\theta^a_k = \theta^a(t_k)$, and $P^jm^j_k = P(t_k)$. Now if we define,
	\begin{align}
		Z^j(t) \eqdef \frac{C^j_{th}}{\eta^j}(\theta^j(t) - \theta^j_{\textSet})
	\end{align}
	then this quantity evolves as,
	\begin{align}
		\dot{Z}^j(t) = -a^jZ^j(t) - \tilde{P}^j(t), \quad \tilde{P}^j(t) = P(t) - \hat{P}^{j,b}(t),
	\end{align}
	where $a^j = R^j_{th}C_{th}^j$, and
	\begin{align}
		\hat{P}^{j,b}(t) = \frac{\theta^a(t) - \theta^j_{\textSet}}{\eta^j R^j_{th}}.
	\end{align}
	Now taking the Laplace transform of both the continuous time odes we have,
	\begin{align}
		Z(s) = -\frac{1}{s+\alpha}\tilde{Y}(s), \ \text{and} \ 
		Z^j(s) = -\frac{1}{s+a^j}\tilde{P}^j(s).
	\end{align}
	Where we have assumed $Z(0) = 0$ and used $\theta^j(0) = \theta_{\textSet}$, so that $Z^j(0) = 0$.
	
	Now, by their respective definitions we have that $\tilde{Y}(s) = \sum_{j=1}^{N}\tilde{P}^j(s)$ so that,
	\begin{align}
		Z(s) &= \sum_{j=1}^{N}-\frac{1}{s+\alpha}\tilde{P}^j(s), \\
			 &= \sum_{j=1}^{N}\frac{s+a^j}{s+\alpha}\frac{-1}{s+a^j}\tilde{P}^j(s) \\ \label{eq:toTakeInvLap}
			 &=\sum_{j=1}^{N}\frac{s+a^j}{s+\alpha}Z^j(s).
	\end{align}
	Now taking the inverse Laplace transform of the equation~\eqref{eq:toTakeInvLap} and applying the bound $\|y(t)\|_\infty \leq \|h(t)\|_1\|u(t)\|_\infty$ for the inverse transforms of the relation $Y(s) = H(s)U(s)$ we have, 
	\begin{align}
		\|Z(t)\|_\infty \leq \sum_{j=1}^{N}\left(1 + \left|1-\frac{R^j_{th}C^j_{th}}{\alpha}\right|\right)\|Z^j(t)\|_\infty.
	\end{align}
	Since the above is valid for any $t\in \mathbb{R}$, we evaluate it at the point $t_k$ to get,
	\begin{align}
		\|Z_k\|_\infty \leq \sum_{j=1}^{N}\left(1 + \left|1-\frac{R^j_{th}C^j_{th}}{\alpha}\right|\right)\|Z^j_k\|_\infty,
	\end{align}
	which is valid for any sequence of times $\{t_k\}_k$ that satisfy $t_k = t_{k-1} + T_s$ with $t_0 = 0$. Now, by assumption in the Lemma the quantity $\|Z^j_k\|_\infty \leq \bar{C}^j$ so that
	\begin{align}
		\left|Z_k\right| \leq \sum_{j=1}^{N}\left(1 + \left|1-\frac{R^j_{th}C^j_{th}}{\alpha}\right|\right)\bar{C}^j, \quad \forall \ k,
	\end{align}
	which is the desired result.$\qed$

Note that Lemma~\ref{lem:hetEnsemModel} here appears in~\cite{hao_aggregate:2015} in continuous time. In our proof we use a connection to continuous time, and the fact that our recursion~\eqref{eq:aggEngDevDyn} is an exact discretization of a certain ode. Additionally, in~\cite{hao_aggregate:2015} the result in Lemma~\ref{lem:hetEnsemModel} is done for a time invariant ambient temperature. As we see from the proof here, the ambient temperature can be time varying and this will not effect the result.

\subsection{ Proof of Lemma~\ref{lem:optProb}}
%To show that the constraint set $\bar{\Omega} = \bigcap_{k}^{N_t-1}\Omega_k$ is non-empty, we start with the baseline trajectory $r_k = 0, \ z_k = 0$ and show that this would allow for a $\bar{\psi}_k \in \Omega_k$ for all $k$. If $r_k$ is zero for all $k$ then $n_k^{\textOn} = \bar{n}^{\textOn}$ constant, which due to~\eqref{eq:switchFracRelWithU} implies that $u_{k-1}^{\textOn}[1] = u_{k-1}^{\textOff}[1]$ for all $k$. Consequently, this implies that  $u_{k-1}^{\textOn} = u_{k-1}^{\textOff} = \bar{u}_{k-1}$ for all $k$ and that $\stuckOnK = \stuckOffK = \bar{\gamma}_k$ for all $k$. Since the inequality constraints are always feasible, then for all $k$ and for all $\bar{n}^{\textOn}$ there exists an element $\bar{\psi}_k = [0,0,\bar{u}_k,\bar{u}_k,\bar{\gamma}_k,\bar{\gamma}_k ]$  such that $\bar{\psi}_k \in \Omega_k$. Thus the set $\bar{\Omega}$ is non empty. 

To show convexity, we use the fact that the intersection of a finite number of convex sets is convex. Each constraint in $\Omega_{t}^{t+H_p -1}$ is convex as the inequality constraints are convex sets and the equality constraints are affine. Thus, $\Omega_{t}^{t+H_p -1}$ is convex as it is the finite intersection of convex sets. 

To show feasibility consider the baseline scenario. In this scenario $Y_k \equiv 0$, which together with the initial condition $Z_t=0$ produces $Z_k \equiv 0$. Hence constraints~\eqref{eq:battModel-in-Omega} and~\eqref{eq:aggZeroEng} are satisfied. From the constraint~\eqref{eq:nkon-in-Omega} we have that $n_k^{\textOn}$ will equal
\begin{align}
	n_k^{\textOn} = \bar{\theta}^a_k - \sum_{j=1}^{N}\frac{\theta_{\textSet}^j}{\eta^j R^j_{th}}\bigg(\sum_{j=1}^{N}P^j\bigg)^{-1}	= \bar{\theta}^a_k - \Gamma,
\end{align}
and the difference satisfies $n_k^{\textOn} - n_{k-1}^{\textOn} = \bar{\theta}^a_k - \bar{\theta}^a_{k-1}$. The  constraint~\eqref{eq:nkon-dyn-in-Omega} is satisfied by
\begin{align}
	\flipOff_k &= \max\{\bar{\theta}^a_{k-1}- \bar{\theta}^a_k,0 \}, \quad \text{and} \\
	\flipOn_k &= \max\{\bar{\theta}^a_{k}- \bar{\theta}^a_{k-1},0 \},
\end{align}
by definition. Upon substituting these choices in the constraints~\eqref{eq:skon-dyn-in-Omega} and~\eqref{eq:skoff-dyn-in-Omega} and using the initial conditions, we have
\begin{align} \label{eq:appendix-stuck-on}
	\stuckOn_{k} & = \sum_{s = k - \tauBA+1}^{k}\max\{\bar{\theta}^a_{s} -\bar{\theta}^a_{s-1},0 \} = \Theta^-_{k}(\tauBA)  \\ \nonumber
	\stuckOff_{k} &=  \sum_{s = k - \tauBA+1}^{k}\max\{\bar{\theta}^a_{s-1} -\bar{\theta}^a_{s},0 \} =  \Theta^+_{k}(\tauBA)
\end{align}
so that by hypothesis, we have
\begin{align}
	\stuckOn_{k} + \Gamma \leq \bar{\theta}^a_{k+1} \leq 1 - \stuckOff_{k} + \Gamma,
\end{align}
which implies that
\begin{align}
\stuckOn_{k} \leq n^{\textOn}_{k+1} \leq 1 - \stuckOff_{k},
\end{align}
and hence the constraint~\eqref{eq:nkon-bound-in-Omega} is satisfied. Additionally, by construction $n_k^{\textOn}$ satisfies~\eqref{eq:ineq-cons-in-Omega} and since $\flipOff_k$ and $\flipOn_k$ are the positive difference of successive values of $n_k^{\textOn}$, they too will satisfy~\eqref{eq:ineq-cons-in-Omega}. By construction $\stuckOffK$ and $\stuckOnK$ are non-negative. Further from the constraint~\eqref{eq:nkon-bound-in-Omega} holding we have $\stuckOnK \leq 1$. Since the fraction of loads stuck on and off satisfy $\stuckOffK + \stuckOnK \leq 1$ we have that $\stuckOffK \leq 1$. Hence, both $\stuckOnK$ and $\stuckOffK$ satisfy~\eqref{eq:ineq-cons-in-Omega}.
   
The above argument, for all of the above constraints, works for any starting index $t$ and any positive planning horizon $H_p$.
$\qed$ 

\subsection{VES constraint}
The BA requires the constraint~\eqref{eq:aggZeroEng} to ensure that the collection of TCLs do not act as generators. We repeat this constraint here for $t=0$:
\begin{align} \label{eq:VESconst}
\sum_{k=0}^{H_p}Y_k = 0.
\end{align}
We now show that this constraint is a necessary condition for the individual TCLs energy constraint~\eqref{eq:invEng}. We assume that $H_p = H_b$, which loses no generality as $H_p$ is arbitrary and would already be a function of $H_b$. Summing~\eqref{eq:invEng} over the $j$ index and expanding the absolute value,
\begin{align}
-\sum_{j=1}^{N}\tilde{E}^j \leq T_s\sum_{j=1}^{N}\sum_{k=0}^{H_p}(m_k^jP - \hat{P}^j_k) \leq \sum_{j=1}^{N}\tilde{E}^j.
\end{align}
%Since the summations are finite and addition of real numbers is both commutative and associative the summation order can be interchanged,
\begin{align} \nonumber
&\implies-\sum_{j=1}^{N}\tilde{E}^j \leq T_s\sum_{k=0}^{H_p}\sum_{j=1}^{N}(m_k^jP - \hat{P}^j_k) \leq \sum_{j=1}^{N}\tilde{E}^j,	\\ \label{eq:nonSymAggEng}
&\implies -\sum_{j=1}^{N}\tilde{E}^j \leq T_s\sum_{k=0}^{H_p}Y_k \leq \sum_{j=1}^{N}\tilde{E}^j.
\end{align}
Converting back to absolute value, the aggregated version of~\eqref{eq:invEng} is
\begin{align} \label{eq:aggTotEng}
T_s\left|\sum_{k=0}^{H_p}Y_k\right| \leq \sum_{j=1}^{N}\tilde{E}^j,
\end{align}
which due to~\eqref{eq:VESconst} will be true for all values of $\tilde{E}^j$, as the RHS term in~\eqref{eq:aggTotEng} is defined to be greater than or equal to zero. If~\eqref{eq:aggTotEng} is not satisfied, then it can be shown through the law of the contrapositive that there would exist at least a single TCL that does not satisfy~\eqref{eq:invEng}. 
In the scenario that the individual TCLs do not have symmetric energy constraints, then the aggregate version of~\eqref{eq:invEng} would resemble~\eqref{eq:nonSymAggEng}; The constraint~\eqref{eq:VESconst} still enforces this. 

\fi

\end{document}